\documentclass[useAMS,usenatbib,twocolumn]{mn2e} \usepackage{times}

\usepackage{graphicx,color}
\usepackage{amsmath,amssymb,color,siunitx}

\input epsf

\usepackage{wrapfig}

\newcommand{\beq}{\begin{equation}} \newcommand{\beqa}{\begin{eqnarray}}
\newcommand{\eeq}{\end{equation}} \newcommand{\eeqa}{\end{eqnarray}}
 
  \newcommand{\gsim}{\ga}
 \newcommand{\vect}[1]{\mbox{\boldmath${#1}$}}
\newcommand{\lmk}{\left(} \newcommand{\rmk}{\right)} 

\newcommand{\lkk}{\left[} \newcommand{\rkk}{\right]} \newcommand{\lla}{\left\langle}
 \newcommand{\rra}{\right\rangle} \newcommand{\so}{M_\odot}
    
  \newcommand{\ven}{\vect n}

\title{Prospects of the local Hubble parameter measurement using gravitational waves from double neutron stars } 

\begin{document}

\author[]
{Naoki Seto$^1$ and Koutarou Kyutoku$^{2,3,4,5}$\\
$^1$Department of Physics, Kyoto University, Kyoto 606-8502, Japan\\
$^2$Theory Center, Institute of Particle and Nuclear Studies, KEK,
Tsukuba 305-0801, Japan\\
$^3$Department of Particle and Nuclear Physics, the Graduate University
for Advanced Studies (Sokendai), Tsukuba 305-0801, Japan\\
$^4$Interdisciplinary Theoretical Science (iTHES) Research Group, RIKEN,
Wako, Saitama 351-0198, Japan\\
$^5$Center for Gravitational Physics, Yukawa Institute for Theoretical
Physics, Kyoto University, Kyoto 606-8502, Japan
}

\maketitle

\begin{abstract} 
Following the detection of the GW170817 signal and its associated
 electromagnetic  emissions, we discuss the prospects of the local Hubble
 parameter measurement using double neutron stars (DNSs). The kilonova
 emissions of GW170817 are genuinely unique in
 terms of the rapid evolution of color and magnitude and we
 expect that,  for a good fraction $\gsim 50\%$ of the DNS events within
 $\sim 200$Mpc, we could identify their host galaxies,  using their
 kilonovae. At present,  the estimated DNS merger rate
 $(1.5^{+3.2}_{-1.2})\times 10^{-6} {\rm Mpc^{-3} yr^{-1}}$  has a large
 uncertainty. But, if it is at the high end, we could measure the local
 Hubble parameter $H_L$ with the level of $\Delta H_L/H_L\sim 0.042$
 ($1\sigma$ level), after the third observational run (O3). This
 accuracy is four times better  than that obtained from GW170817 alone,
 and we will be able to examine the Hubble tension at $2.1\sigma$
 level. 
\end{abstract}



\section{introduction}

The gravitational wave (GW) signal GW170817  was detected  at the  signal-to-noise ratio (SNR) of 32.4 that is the largest value among the  GW signals detected so far (Abbott et al. 2017a).  From the estimated masses, the signal is considered to be generated by a DNS inspiral.   After the GW detection, the associated electromagnetic (EM) emissions were discovered worldwide at various wavelengths (Abbott et al. 2017b). These sequential events brought profound impacts broadly on astronomical and physical communities.    Here, in the face of the current torrent of research papers, we do not mention general aspects of GW170817, but rather concentrate on our main topic, observational cosmology.

It has been long known that, using GWs from binary inspirals (often called the standard sirens), we can estimate the distance to the source, solely based on the first principle of physics (Schutz 1986, Krolak \& Schutz 1987). This shows a remarkable contrast to the traditional distance ladder that relies  heavily  on  various empirical relations. 
Meanwhile, because of the simple scaling property of general relativity, it is not straightforward to estimate the redshift of the binary only from GWs (see also Chernoff \& Finn 1993, MacLeod \& Hogan 2008, Messenger \& Read 2012).  Therefore, in order to utilize the standard sirens efficiently, it would be crucially advantageous,  if we can identify the transient EM signals associated with the GW events (see {\it e.g.} Holz \& Hughes 2005, Nissanke et al. 2010). But, we had been far from confident whether such multi-messenger observations actually work. 

Now, this concern is largely untangled by the  followup observations of
GW170817 and the resulting  identification of its host galaxy NGC4993 at
$z=0.010$ (after the peculiar velocity correction, Abbott et al. 2017c, see also  Hjorth et al. 2017). In fact, its kilonova (also called the macronova) emission turned out to be genuinely unique in
 terms of the rapid evolution of color and magnitude, also showing a characteristic time
profile. It
is true that  we only have the single DNS event and additional ones are
essential to understand the basic properties of the EM
counterparts. But, now, we can expect long-term development of
observational cosmology, by using DNSs as a powerful probe.

In the near future, around the entrance of this new avenue, our primary target would be the Hubble parameter, as already discussed in the pioneering work by Schutz (1986) more than 30 years ago. Indeed, using the distance $\sim 40$Mpc estimated from the GW170817 signal and the redshift of its host galaxy, the LIGO-Virgo team reported the Hubble parameter $H_0=70^{+12}_{-8}~{\rm km~sec^{-1}Mpc^{-1}}$ (Abbott et al. 2017c). Here the error bar represents 68.3\% probability range.

The Hubble parameter is one of the most fundamental cosmological
parameters, since the discovery of the cosmic expansion in 1929. But
this parameter has attracted much attention quite recently.  We have a
9\%  mismatch between the locally estimated value $73.24\pm1.74~{\rm km~sec^{-1}Mpc^{-1}}$ and that determined from the cosmic microwave background $66.93  \pm0.62~{\rm km~sec^{-1}Mpc^{-1}}$ (Riess et al. 2016, Planck Collaboration 2016).  This tension might be caused by an unidentified systematic error in the two types of measurements or might, in fact, imply a challenge to the standard cosmological model.  In any case, the newly established method based on the DNSs could make a notable contribution to the Hubble tension.

In this paper, we discuss the prospects of gravitational-wave
observational cosmology with the forthcoming third observation run  (O3)
of the LIGO-Virgo collaboration (Abbott et al. 2016) and its follow-on operations (see Zhao \& Wen 2017 for the third generation detectors). 
Our results would be also useful  to discuss related topics such as the efforts to suppress the amplitude calibration error of the GW measurement (see {\it e.g.} Vitale et al. 2012, Tuyenbayev et al. 2017, Cahillane et al. 2017) or the   observational strategy for the EM counterpart search (Cowperthwaite et al. 2017).

This paper is organized as follows. In \S2, we discuss the kilonova signals and contaminations at the host galaxy identification, taking into account the actual observational results of GW170817. In \S3, we derive analytical expressions to evaluate the expected number of DNS detections and the averaged distance error at the GW data analysis. Then, in \S4, we discuss the prospects of the Hubble parameter measurement in the near future.  \S5 is a brief summary of this paper. 
Following the standard convention, we assume the DNS masses at  $1.4\so+1.4\so$ whose chirp mass is only $\sim 2\%$  different from that of GW170817. 

\section{kilonova signals for the host galaxy identification}

The loudness of the optical sky always stands in the way to identify
kilonovae in followup observations (Cowperthwaite \& Berger 2015, Tanaka
2016, Cowperthwaite et al. 2017). Here, we argue that a
good fraction $\gtrsim 50\%$ of kilonovae for DNS events within $\sim
\SI{200}{Mpc}$ can be identified by followup observations in optical
(and hopefully near-infrared) bands incorporating insights obtained from
observations of the kilonova associated with GW170817. Once an
electromagnetic counterpart such as the kilonova is successfully
identified, the host galaxy will be determined relatively easily,
because the expected cosmological redshift will at most be
0.05--0.1. Real-time identification is not necessary for our purpose,
i.e., determining the host galaxy, and accordingly we do not worry about
the lack of template images (Cowperthwaite \& Berger 2015).

The key findings from the kilonova associated with GW170817 are
summarized as follows (see, e.g., McCully et al. 2017, Shappee et
al. 2017, Siebert et al. 2017, Utsumi et al. 2017). The emission becomes
bright right after the merger , say $\sim \SI{1}{day}$, peaking in blue
optical bands (Shappee et al. 2017). While the magnitude in optical
bands such as the \textit{g}-band drops very rapidly (Siebert et
al. 2017), near-infrared bands sustain bright emission for a few to
several days, where longer emission is found in longer wavelengths
(Utsumi et al. 2017). Even in the relatively red \textit{z}-band, the
emission becomes dim by \SI{2.5}{mag} in only 6 days (Utsumi et
al. 2017).  The corresponding
change in the peak wavelength makes the color evolution exceedingly
rapid (McCully et al. 2017). The rapid decline of magnitudes and the
rapid reddening are both distinctive features of the kilonova (see also
Cowperthwaite \& Berger 2015, Cowperthwaite et al. 2017). Furthermore,
the spectrum is mostly featureless and becomes red as early as a few
days after the merger (McCully et al. 2017, Siebert et al. 2017). Such
spectra are not observed for other known transients, and thus make the
kilonova very unique (Shappee et al. 2017). It should be worth noting
that GW170817 appears to be observed from relatively polar directions
(Abbott et al. 2017a).

Turning now to identification of future kilonovae. Fast optical
transients as significant contaminants are summarized comprehensively in
Cowperthwaite \& Berger (2015). Among the fast transients, the type Ia
supernova outnumbers the kilonova at given brightness. Fortunately, they
will be easily distinguished due to their significantly slow time
evolution compared to that of kilonovae (see also Scolnic et
al. 2017). Elimination of type Ia supernovae could further benefit from
the line structures in the spectrum if it is taken and the presumably
high redshift. On the other extreme, stellar flares last less than a day
and will be eliminated by requiring detections in multiple
nights. Quiescent emission of the underlying star could be detected
later for further secure elimination.

Taking the estimated rate of fast transients, we expect that the type
.Ia supernova and so-called Pan-STARRS fast transients can serve as
significant contaminants (Cowperthwaite \& Berger 2015). A remarkable
point is that the peak brightness of the kilonova associated with
GW170817 was found to be brighter by $\gtrsim \SI{1.5}{mag}$ (Utsumi et al. 2017), or
equivalently by a factor of $\gtrsim 4$ than the model adopted in
Cowperthwaite \& Berger (2015). This means that the number of fast transients that can
compete with kilonovae will be reduced by a factor of 8. Thus, the
number of type .Ia supernovae and Pan-STARRS fast transients will only
be $\approx 0.2$ and $\lesssim 1$, respectively, even for
\SI{100}{deg^2} sky localization. Furthermore, the rapid decline of the
kilonova associated with GW170817 is hardly reproduced by Pan-STARRS
fast transients. Thus, a large fraction of Pan-STARRS fast transients,
say $\gtrsim 80\%$, would be removed by requiring a moderate decline
rate that does not significantly remove blue kilonovae. Overall, we
expect that more than $\gtrsim 50\%$ of blue kilonovae can be identified
based on the brightness and decline rates.

The rapid reddening and/or red color, once detected, will serve as a
powerful tool to distinguish kilonovae from other transients as
previously thought (Cowperthwaite \& Berger 2015, Tanaka 2016). Note
that the blueness of the kilonova associated with GW170817 does not
exclude existence of red kilonovae. Particularly, we still expect to
observe kilonovae without blue components for edge-on binaries, for
which lanthanide-enriched dynamical ejecta should be obscuring any blue
emission (Kasen, Fern{\'a}ndez \& Metzger 2015). A substantial fraction
of red kilonovae can be securely identified as electromagnetic
counterparts to DNS events further aided by rapid magnitude
evolution. Quantitatively, requiring $i-z \gtrsim \SI{0.4}{mag}$ will
remove most of the contaminants, say $\gtrsim 90\%$, without discarding
red kilonovae significantly (Cowperthwaite \& Berger 2015). The
requirement may be loosened to $\gtrsim \SI{0}{mag}$ when we
additionally require the rise or decline time to be $\lesssim
\SI{4}{day}$, but about a half of kilonovae may be lost for the cut
based on the decline (Cowperthwaite \& Berger 2015).

To summarize, we expect that more than 50\% of kilonovae may be
identified successfully by first seeking rapid blue components, and next
red components. In any case, properties of kilonovae discussed here rely
heavily on a single event GW170817 combined with theoretical knowledge,
and future detections of a variety of kilonovae in followup observations
to DNS events are crucially important to refine the selection
criteria. The fraction of identifiable kilonovae may be increased in the
near future by understanding their characteristics with actual
observations.

\section{signal detection and analysis}

In this section, we discuss analytical expressions for the detectable volume of DNSs and the appropriately averaged  their distance estimation errors. We basically follow the formulation in  Cutler \& Flanagan (1994). In addition, using the method developed in Seto (2015), we perturbatively include the geometrical information of the detector network.  We also count the differences in sensitivities among  detectors, and derive  expressions convenient for  statistical study of the local Hubble parameter measurement.

\subsection{detectable volume}

The SNR $\rho_i$ of a detector $i$ can be written as
\beq
\rho_i^2=10 \lmk\frac{d_{h,i}}D  \rmk^2 [F_{+,i}(\ven,\psi)^2 \frac{(1+v^2)}4+F_{\times,i}(\ven,\psi)^2v^2] ,\label{ri}
\eeq
where $F_{+,i}$ and $F_{\times,i}$ are the beam pattern functions that depend on the source direction $\ven$ and the polarization angle $\psi$ (Thorne 1987). The quantity $d_{h,i}$ is the horizon distance for $\rho_i=10$. 
We have the relation $d_{h,i}=2.26 d_{r,i}$ with the detection range
$d_{r,i}$ for the same SNR (see {\it e.g.} Chen et al. 2017). In
eq.(\ref{ri}), $D$ is the (luminosity) distance to the binary and 
$v$ is the cosine of its inclination angle with $|v|=1$ for face-on and $v=0$ for edge-on. 

If the detector noises are uncorrelated, the total SNR  $\rho$ of a detector network is given by 
\beq
\rho^2=\sum_i \rho_i^2,
\eeq
Following Cutler \& Flanagan (1994), we express $\rho^2$ in the form
\beq
\rho^2=\frac{\sigma(\ven)}{D^2} [c_0(v)+\epsilon (\ven)c_1(v)\cos4\psi] \label{rho}
\eeq
with $c_0(v)\equiv (1+v^2)^2/4+v^2$ and $c_1(v)\equiv (1+v^2)^2/4-v^2$.
Here the function $\sigma (\ven)$ shows the total sensitivity of the network to  GWs coming from the direction $\ven$. Meanwhile, the function $\epsilon(\ven)$ represents the asymmetry of the sensitivities to two (appropriately decomposed) orthogonal polarization modes.  We generally have $0\le \epsilon (\ven)\le 1$. If the network is blind to one of the modes, we have $\epsilon(\ven) =1$.

From eq.(\ref{rho}), for a given SNR threshold $\rho_T$, the maximum detectable distance $D_{max}$  is given by
\beq
D_{max}=\frac{1}{\rho_T} \sigma(\ven)^{1/2} [c_0(v)+\epsilon(\ven) c_1(v)\cos4\psi]^{1/2},\label{Dmax}
\eeq
and can be regarded as a function of the four angular parameters $(\ven,\psi,v)$.
 
To simplify expressions below,  we introduced the averaging operation with  the angular parameters;
\beq
\int dA [\cdots]\equiv \frac1{4\pi}\int_{4\pi} d\ven~\frac1{2\pi}\int_0^{2\pi}d\psi~ \frac12\int_{-1}^1  dv[\cdots].
\eeq
Then the effective volume $V$ for the detection threshold $\rho_T$ is given as
\beq
V=\int dA \int_0^{D_{max}} dD 4\pi D^2= \rho_T^{-3}U \label{vol},
\eeq
where we defined
\beq
U\equiv\frac{4\pi}3 \rho_T^3 \int dA D_{max}^3.
\eeq
Using eq.(\ref{Dmax}), we formally have 
\beq
U=\frac{4\pi}3\int dA \sigma(\ven)^{3/2}  [c_0(v)+\epsilon(\ven) c_1(v)\cos4\psi]^{3/2}.
\eeq
We now evaluate this expression. 
Note that, because of the power 3/2, the four-dimensional integrals $dA$ cannot be performed separately (Seto 2015).  But we can overcome this difficulty  by perturbatively  expanding the term proportional to $\epsilon(\ven)$ as follows
\beqa
U&=&\frac{4\pi}3\int dA \sigma(\ven)^{3/2}  c_0(v)^{3/2} \Big[ 1+\frac32\frac{\epsilon(\ven) c_1(v)\cos4\psi}{c_0(v)}\nonumber\\
& & +\frac38\lmk \frac{\epsilon(\ven) c_1(v)\cos4\psi}{c_0(v)}\rmk^2+\cdots\Big].
\eeqa
After performing integrals separately, we obtain 
\beqa
U&=&\frac{4\pi}3g\times 0.821(1+0.01s_2+2.1\times 10^{-4}s_4+\cdots)\label{vb}.
\eeqa
Here we used numerical results such as
\beq
\frac12\int_{-1}^1 dv c_0(v)^{3/2}=0.821,
\eeq
and also  defined 
\beqa
g&\equiv& \frac1{4\pi} \int_{4\pi} d\ven \sigma(\ven)^{3/2},\\
s_m&\equiv& \frac1{4\pi g} \int_{4\pi} d\ven \sigma(\ven)^{3/2} \epsilon(\ven)^m
\eeqa
for even $m$.
For the quantity $U$, all the geometrical information of the network are contained in  $g$ and $s_m$. 

Since $0\le \epsilon(\ven) \le 1$, we have the following  inequalities for the integrals $s_m$
\beq
0\le \cdots\le s_4\le s_2\le 1.
\eeq
Therefore, after dropping the corrections $\propto s_m$ in eq.(\ref{vb}), we get  a good approximation
\beq
U\simeq 3.44 g \label{ap1}
\eeq 
with  relative error less than 1\% (Seto 2015, see also Schutz 2011).

For a network with  a single detector $i$, we identically have $\epsilon(\ven)=1$ and  $s_m=1$. We also have
\beq
V=\frac{4\pi}3 \lmk\frac{10}{\rho_i}  \rmk^3 d_{r,i}^3 
\eeq
because of  the definition of the detection range $d_{r,i}$.

\subsection{distance error}
We assume that the source direction $\ven$ is accurately determined by
the sky position of the EM counterparts such as the kilonova. Then from the information related to the extrinsic properties of GWs,  we  need to simultaneously estimate just the four extrinsic parameters, $D$, $\psi$,  $v$ and the initial phase of the wave. 
From the Fisher matrix of these parameters (Cutler \& Flanagan 1994), the variance of the relative distance error for a binary is given by 
\beq
\lla \lmk\frac{\Delta D}{D}\rmk^2\rra={4D^2} \frac{(1+v^2)-\epsilon(\ven)(1-v^2)\cos4\psi}{\sigma(\ven) (1-\epsilon(\ven)^2) (1-v^2)^2}.\label{dd}
\eeq
This expression depends on the four angular parameters $(\ven,\psi,v)$ as well as the distance $D$. Due to the singularity of the Fisher matrix, this expression diverges at $|v|\to 1$ (face-on) and overestimates the variance, compared with a more elaborate nonlinear estimation (see {\it e.g.} Nissanke et al. 2013, Rodriguez et al. 2014). In contrast, for the edge-on binaries, eq.(\ref{dd}) would be an underestimation, especially for low SNRs.

Assuming a homogeneous and isotropic binary distribution, we can derive
an averaged error in the relative distance $\sigma_{\rm lnD}$ for
binaries with $\rho>\rho_T$ as
\beq
\sigma_{\rm lnD}^2= 
\lkk  \frac{\int dA \int_0^{D_{max}}dD\lla \lmk\frac{\Delta D}{D}\rmk^{2}\rra^{-1}4\pi  D^2 }{\int dA \int_0^{D_{max}}dD4\pi D^2  }           \rkk^{-1}\label{int}.
\eeq
After the $dD$ integral, we have 
\beq
\sigma_{\rm lnD}^2=\rho_T^{-2} \frac{U}{X} \label{dd}
\eeq
where we defined 
\beqa
X&\equiv &\pi  \int dA   \sigma(\ven) (1-\epsilon^2) (1-v^2)^2  \nonumber\\
& &\times [(1+v^2)-\epsilon(\ven)(1-v^2)\cos4\psi]^{-1} \int_0^{D_{max}} dD\label{x1}\\
&=&\pi  \int dA \sigma(\ven)^{3/2}  (1-v^2)^2  [c_0(v)+\epsilon(\ven) c_1(v)\cos4\psi]^{1/2} \nonumber\\
& & \times(1-\epsilon^2)[(1+v^2)-\epsilon(\ven)(1-v^2)\cos4\psi]^{-1}.
\eeqa
As shown in the $dD$ integral in eq.(\ref{x1}), $X$ is not dominated by
small $D(\ll D_{max}) $. Therefore, for a sufficiently large number of
DNS events, the statistical fluctuations of our estimation $\sigma_{\rm
lnD}$ would be small.

In the same manner as $U$ in the previous subsection, after expanding the relevant factors in  $X$ and  averaging with  the four angular parameters, we get
\beq
X=g(0.966-0.574s_2-0.158s_4-0.068s_6-0.037s_8-\cdots).
\eeq
This expression is our new result  and would be useful for statistical discussion on the local Hubble parameter measurement.

\section{Prospects of O3 and beyond}

Based on the expressions derived in the previous section, we now discuss the prospects of the forthcoming observational runs.

\subsection{observation plan}

In Table \ref{tab1}, we summarize the actual results of the past two runs, O1 and O2, and the planned parameters for the future runs  O3 and O4. Here we denote the 2020+run (in Abbott et al. 2016) simply by O4.  In Table \ref{tab1}, the duration  $T_{d}$ for O1 and O2 are the total time for the simultaneous operation of the two LIGO detectors (based on Abbott et al. 2017a). For O3 and O4, the observational time relevant for the Hubble parameter estimation should be
\beq
T_{obs}=f_d T_{d}
\eeq
with the time  fraction $f_d$ for the simultaneous operation of all the three detectors. The duration $T_{d}$ of O4 is not explicitly presented in Abbott et al. (2016). 

In Table 1, the detection ranges $d_{r,i}$ are given for the threshold $\rho_T=10$ (in contrast to the conventional $\rho_T=8$). In the 6th column, we present the four-dimensional volume $VT_{d}$ using eq.(\ref{vol}) for $\rho_T=10$.
We also present  $s_2$, $s_4$, and $\sqrt{U/X}$ required  for the estimation of the relative distance error $\sigma_{\rm lnD}$.

At the stage 2024+ (Abbott et al .2016), KAGRA is planned to  join the detector network with $d_{r,i}=112$Mpc, in addition of two LIGOs ($d_{r,i}=152$Mpc) and Virgo ($d_{r,i}=100$Mpc).  For these four detectors, we have $V=68\times 10^6{\rm Mpc^3}$ and  $\sqrt{U/X}=2.5$.

The  DNS merger rate $R$ estimated after the GW170817 detection is  (Abbott et al. 2017a)
\beq
R=(1.5^{+3.2}_{-1.2})\times 10^{-6} {\rm Mpc^{-3} yr^{-1}}
\eeq
(90\% probability range).
We hereafter denote $R=f_R R_0$ with the median value $R_0=1.5\times 10^{-6} {\rm Mpc^{-3} yr^{-1}}$ and the scaling parameter $f_R$.
Then the expected DNS events is given by
\beq
N=R V (f_d T_{d})=(f_R f_d) R_0 T_{d}  U \rho_T^{-3}.
\eeq 
In reality, we will  be able to identify the host galaxies for not all of  the $N$ events. Therefore, we introduce the probability $f_{E}$ for the successful host galaxy identification, and  
use the total DNS events $N_{E}=f_{E}N$ for estimation of the local Hubble parameter.
Here, for simplicity, we neglect the dependence of $f_E$ on the distance $D$ and the inclination $v$. 
 Using Table 1, we explicitly have
\beq
N_E=A(f_R f_d f_E)  \lmk \frac{T_{d}}{\rm 1yr}\rmk  \lmk   \frac{\rho_T}{10}\rmk^{-3} \label{ne0}
\eeq
with the numerical coefficients $A=32$ for O3 and 69 for O4.
In Table \ref{tab2}, we summarize the parameters that appear in this expression and are also useful for discussions below.

\begin{table*}
  \begin{tabular}{|l|c|ccc|c|ccc|}\hline
     &  duration $T_{d}$ & $d_{r,i}$: LIGO-H & $d_{r,i}$: LIGO-L &$d_{r,i}$: Virgo & $VT_{d}[10^6{\rm Mpc^3 yr}]$ & $s_2$ & $s_4$ & $\sqrt{U/X}$ \\ \hline
  O1 & 0.14yr & 56Mpc & 56Mpc & & $0.29^*$ &  & &\\
  O2  & 0.3yr$^*$ & 38Mpc & 77Mpc & 21Mpc & $0.8^*$ & $0.859$  &0.769 &3.1 \\
  O3 & 1yr &116Mpc & 116Mpc & 60Mpc & 21 &0.756 & 0.626 &2.8 \\
  O4 (2020+) & &152Mpc & 152Mpc & 72Mpc &46 &0.776& 0.651 &2.9 \\
  \hline
  \end{tabular}
\caption{Parameters for each observational run. We denote 2020+ observation (Abbott et al. 2016) by O4. Detection ranges $d_{r,i}$ for each detector  is given for  $1.4\so+1.4\so$ DNS and the threshold SNR=10 (from Abbott et al. 2016, but O2 from Abbott et al. 2017a).   The four-dimensional volume $VT_{d}$ for O4 is given for $T_{d}=1$yr. The numbers with the asterisk are calculated  without Virgo.  }
\label{tab1}
\end{table*}

\subsection{local Hubble parameter measurement}

Next, we discuss the error in estimation of the local Hubble parameter
 $H_L$.  For each DNS (label $j=1,\cdots,N_E$) with identified host galaxies, 
 we can estimate the Hubble parameter 
\beq
H_j=\frac{cz_j}{D_j},
\eeq 
using the measured redshift $z_j$ of the host galaxy and the distance $D_j$ from GW data analysis.  But both of them contain errors $\delta z_j$ and $\delta D_j$.  The former would be dominated by the local peculiar velocity $ v_j$ as
\beq
\delta z_j\sim v_j/c,
\eeq
and the latter $\delta D_j$ would be the parameter estimation error at GW data analysis.
Then we have
\beq
\frac{\delta H_j}{H_L}\simeq \frac{v_j}{cz_j}+\frac{\delta D_j}{D_j}.
\eeq 
From eq.(\ref{dd}) and Table \ref{tab1}, the magnitude of the relative distance error is roughly estimated as
\beq
\frac{\delta D_j}{D}\sim \frac1{\rho_T} \sqrt{\frac{U}{X}}\sim 0.3 \lmk \frac{\rho_T}{10}  \rmk^{-1} .
\eeq
Meanwhile, given the typical one-dimensional velocity of galaxies $\sim 400\, {\rm km\,sec^{-1}}$ (Strauss \& Willick 1995), we have ${v_j}/{cz_j}\sim 0.05$ for DNS distance $D_j\sim 100$ Mpc ($cz_j\sim 7000{\rm km~sec^{-1}}$). 
Therefore, for individual DNS events, the error for the Hubble parameter is approximately given by 
\beq
\frac{\delta H_j}{H_L}\simeq\frac{\delta D_j}{D_j}.
\eeq
Statistically using totally $N_{E}$ DNSs, we have the estimation error for the local Hubble parameter
\beq
\frac{\Delta H_L}{H_L}\sim f_F\frac{\sigma_{\rm lnD}}{\sqrt N_{E}} \simeq  \sqrt{\frac{U}{X}} \frac{f_F\rho_T^{1/2}}{(f_d f_R f_{E}R_0 T_{d})^{1/2} }. \label{fh}
\eeq

As mentioned earlier, the original expression (\ref{dd}) based on the Fisher matrix could be both over- and underestimate the variance, compared with a more elaborate nonlinear analysis. To include these mismatches, we introduced the correction factor $f_F$ in eq.(\ref{fh}).  We should have $f_F\to 1$ in the limit $\rho_T\to \infty$.

\begin{table}
  \begin{tabular}{ll}\\\hline
$f_d$  &  fraction of time with triple detector operation\\
$f_R$  &  the factor for the local DNS merger rate \\
$f_{E}$  & the fraction of DNS events with identified host galaxy\\
$f_{F}$  & the correction factor for the Fisher matrix analysis\\
$\rho_T$  & SNR threshold for GW signal\\
$T_{d}$ &   total observation duration \\\hline
 \end{tabular}
\caption{ Basic parameters for statistical analysis of  multiple events. }
\label{tab2}
\end{table}

\subsection{case study for O3 and beyond }

Using expressions in the previous subsections and  assuming the planned
detector sensitivity in Table 1, we have eq.~\eqref{ne0} for the
expected DNS events with identified host galaxies, and 
\beq
{\Delta H_L}/H_L\sim B f_F (f_R f_d f_{E})^{-1/2} (\rho_T/10)^{1/2}(T_{d}/1{\rm yr})^{-1/2} \label{dh1}
\eeq
for the error of the local Hubble parameter. Here we have the numerical factors $(A,B)=(32,0.049)$ for O3 and (69,0.035) for O4.
These expressions still depend on the six parameters $(f_R, f_d, f_E, f_F, T_{d},\rho_T)$. 
 Therefore, we evaluate eqs. (\ref{ne0}) and (\ref{dh1}) for the three cases below.
We commonly assume $T_{d}=1$yr for O3 and 2yr for O4, and also fix the SNR threshold at  $\rho_T=10$. In reality, at the stage of O4, we can use the results of O3. By combining O3 and O4, the number $N_E$ would be 23\% larger and the error ${\Delta H_L}/H_L$ would be $\sim 11\%$ smaller. But, for simplicity,  we discuss the results O3 and O4 separately. 

 \subsubsection{optimistic case.}
We assume  the higher end of the DNS merger rate $f_R=3.1$ and the high efficiencies  $f_E=f_d=0.8$ with the correction factor $f_F=1.2$. 

At the end of O3, we have $\Delta H_L/H_L\sim 0.042$ with $N_E\sim 63$.
Therefore, if the current mismatch $\sim 9\%$  between the local and global
Hubble parameters  might be examined at $2.1\sigma$ level.  Then, after O4, we will have $\Delta H_L/H_L=0.021$ with $N_E=274$.  

The cosmic variance due to the coherence of the peculiar velocity field is estimated to be $\sim 1\%$ at the survey depth of $D\sim 200$Mpc  (Shi \& Turner 1998) and could become a potential concern at O4.  Even for O3, there would be a strong motivation to suppress the amplitude calibration error much lower than $5\%$ (Tuyenbayev et al. 2017).

\subsubsection{standard case}

We assume the median value $f_R=1.0$ for the merger rate, and the efficiencies $f_E=f_d=0.6$ with the factor $f_E=1.3$. As easily understood from Eqs. (\ref{ne0}) and (\ref{dh1}), the factor $f_R$ is the major cause of the difference from the optimistic case  above.

With O3, we have $\Delta H_L/H_L\sim 0.11$ with $N_E\sim 12$. Among the expected twelve events, the maximum SNR is roughly estimated to be $10\times 12^{1/3}\sim 23$, and smaller than that of GW170817.  The error $\Delta H_L/H_L\sim 0.11$ is not so different form $0.15$ obtained from  GW170817. 
After O4, we have   $\Delta H_L/H_L\sim 0.053$ with $N_E=50$, and  can examine the tension at $2\sigma$ level.

\subsubsection{pessimistic case}

We assume the lower end of the rate $f_R=0.2$, and the efficiencies
$f_E=f_d=0.5$  with $f_F=1.4$.  We have $N_E\sim 2.3$ for O3 and 10 for
O4.  Even with O4, the error $\Delta H_L/H_L$ remains 0.15 and is
comparable to that from GW170817. If the time fraction $f_d$ is less
0.5, the error becomes even larger.   We expect that GW observation is
not likely to  play a critical  role to examine the Hubble tension in
the next five years for this case.

\section{summary}

The GW170817 event clearly demonstrated that the DNSs could become ideal standard sirens accompanied by characteristic EM signals for host galaxy identification (Abbott et al. 2017a). The kilonova of GW170817 is genuinely unique in
 terms of the rapid evolution of color and magnitude and we expect that,  for a good fraction $\gsim 50\%$ of the DNS events within $\sim 200$Mpc, we can identify their host galaxies,  using their kilonovae. Therefore, the DNSs will become a powerful and reliable tool for observational cosmology. 
Our immediate target would be the locally measured Hubble parameter that
currently has a 9\% tension with the value obtained form the cosmic microwave background (Riess et al. 2016, Planck Collaboration 2016). Considering these circumstances, we discussed the prospects of the Hubble parameter measurement using DNSs observed during the forthcoming LIGO-Virgo observational runs (Abbott et al. 2016).

In order to evaluate the measurement error of the Hubble parameter estimated form multiple DNSs, we derived convenient expressions, and applied them for the three observational scenarios. 
If the DNS merger rate is at the high end of the current constraint $\sim 3.7 \times 10^{-6}{\rm Mpc^{-3}yr^{-1}}$ and the planed sensitivities are realized for LIGO and Virgo, we could attain the accuracy $\Delta H_L/H_L\sim 0.042$ with O3.  Then, the Hubble tension might be verified at $2.1\sigma$ or we might indicate a potential systematic error for the traditional cosmoligical distance probes. 
Also, this precision would give a strong motivation to suppress the amplitude calibration errors of the ground-based detectors. On the other hand, if the DNS merger rate is at the low end  $\sim 0.2 \times 10^{-6}{\rm Mpc^{-3}yr^{-1}}$, even with the 2020+ observation,  it would be unlikely to go significantly beyond the level $\Delta H_L/H_L\sim 0.15$ already obtained by GW170817 whose $SNR =32.4$ is contrastingly at the high end tail. 

In any case, additional DNS events with O3 would be indispensable to further constrain the DNS rate and also  better understand the EM counterparts, especially the anisotropies of  kilonova emissions.   

\section*{Acknowledgments}
We would like to than Masaomi Tanaka for helpful comments.
 This work is supported by JSPS Kakenhi Grant-in-Aid for Scientific Research
 (Nos.~15K65075, 16H06342, 17H01131, 17H06358).



\end{document}